\documentclass[12pt]{article}

\usepackage{graphicx}
\begin{document}

\newcommand{\siml}{\stackrel{<}{\sim}}
\newcommand{\simg}{\stackrel{>}{\sim}}
\newcommand{\lleq}{\stackrel{<}{=}}



%
\begin{center}
{\large\bf
Synchrony and variability induced by spatially correlated
additive and multiplicative noise \\
in the coupled Langevin model 
} 
\end{center}

\begin{center}
Hideo Hasegawa
\footnote{hideohasegawa@goo.jp}
\end{center}

\begin{center}
{\it Department of Physics, Tokyo Gakugei University  \\
Koganei, Tokyo 184-8501, Japan}
\end{center}
\begin{center}
({\today})
\end{center}
\thispagestyle{myheadings}

\begin{abstract}
The synchrony and variability have been discussed of
the coupled Langevin model subjected to spatially correlated
additive and multiplicative noise.
We have employed numerical simulations and
the analytical augmented-moment method which is the second-order 
moment method for local and global variables [H. Hasegawa, Phys. Rev. E
{\bf 67}, 041903 (2003)]. It has been shown that the synchrony of an ensemble 
is increased (decreased) by a positive (negative) spatial correlation 
in both additive and multiplicative noise. Although the variability 
for local fluctuations is almost insensitive to spatial correlations, 
that for global fluctuations is increased (decreased) 
by positive (negative) correlations. When a pulse input is applied, 
the synchrony is increased for the correlated multiplicative noise, 
whereas it may be decreased for correlated additive noise coexisting 
with uncorrelated multiplicative noise. An application of our study to neuron 
ensembles has demonstrated the possibility that information is conveyed by 
the variance and synchrony in input signals, which accounts for some neuronal 
experiments.  

\end{abstract}

\noindent
\vspace{0.5cm}

{\it PACS No.}: 05.10.Gg, 05.40.Ca, 84.35.+i
%

 

\newpage

\section{Introduction}

It has been realized that the coupled Langevin model is a valuable 
and useful model for a study of various phenomena observed in stochastic 
ensembles (for a recent review, see Ref. \cite{Lindner04}).
Independent (uncorrelated) additive and/or multiplicative noise
has been widely adopted for theoretical analyses
because of its mathematical simplicity. In natural phenomena, however,
there exist some kinds of correlations in noise, such as
spatial and temporal correlations, and a correlation 
between additive and multiplicative noise. 
In this paper we will pay our attention to the spatial correlation in noise.
The Langevin model has been usually discussed with the
use of the Fokker-Planck equation (FPE) for the probability distribution.
In the case of correlated additive noise only,
the probability distribution is expressed by the multivariate
Gaussian probability with a covariance matrix. 
The effect of correlated additive noise has been extensively studied 
in neuroscience, where it is an important and essential problem to study 
the effect of correlations in noise and signals
(for a review, see Ref. \cite{Averbeck06}).
It has been shown that the synchrony and variability in neuron ensembles
are much influenced by the spatial correlations
\cite{Burkitt99}-\cite{Brown08}.
The spatial correlation in additive noise enhances the synchrony of firings 
in a neuron ensemble, while it works to diminish beneficial roles of 
independent noise, 
such as the stochastic and coherent resonances and the population 
(pooling) effect 
\cite{Averbeck06,Liu01,Wang04,Kwon05};
related discussions being given in Sec. 3.

The problem becomes much difficult when multiplicative noise exists, 
for which the probability distribution generally becomes a non-Gaussian.
Although an analytical expression of the stationary probability distribution 
for uncorrelated multiplicative noise is available, that for correlated 
multiplicative noise has not been obtained yet.
Indeed, only little theoretical study of the effect of 
spatially correlated multiplicative noise has been
reported for subjects such as the noise-induced phase separation \cite{Ibanes99}
and the Fisher information \cite{Abbott99}-\cite{Wu04}, 
as far as the author is concerned.

In a recent paper \cite{Hasegawa07a}, we have studied stationary and
dynamical properties of the coupled Langevin model
subjected to uncorrelated additive and multiplicative noise.
We employed the augmented moment method (AMM)
which was developed for a study of stochastic systems with finite 
populations \cite{Hasegawa03a,Hasegawa06}.
In the AMM, we consider global properties of ensembles, taking account 
of mean and fluctuations (variances) of local and global variables.
Although a calculation of the probability
distribution for the spatially correlated multiplicative 
noise with the use of the FPE is very difficult as mentioned above, 
we may easily study its effects by using the AMM.
It is the purpose of the present paper to apply the AMM to the
coupled Langevin model including spatially correlated multiplicative noise
and to study its effects on the synchrony and variability. 

The paper is organized as follows. In Sec. 2, we discuss the AMM 
for the spatially correlated Langevin model. 
With the use of analytical AMM and numerical methods,
the synchrony and variability of the coupled Langevin model are investigated.
In Sec. 3, previous studies on the correlated multiplicative noise
using the Gaussian approximation
\cite{Abbott99}-\cite{Wu04} are critically discussed.
An application of our study to neuron ensembles
is also presented with model calculations.
The final Sec. 4 is devoted to our conclusion. 

\section{Formulation}

\subsection{Adopted model}

We have assumed the $N$-unit coupled Langevin model subjected to 
spatially correlated additive and multiplicative noise.
The dynamics of a variable $x_i$ ($i=1$ to $N$) is given by
\begin{eqnarray}
\frac{dx_{i}}{dt} &=& F(x_{i}) 
+H(u_{i})
+ G(x_{i}) \: \eta_{i}(t) + \xi_{i}(t), 
\label{eq:A1}
\end{eqnarray}
with
\begin{eqnarray}
u_{i}(t) &=& \left( \frac{w}{Z} \right) 
\sum_{j (\neq i)} \:x_{j}(t) + I_i(t), \label{eq:A0}\\
H(u) &=& \frac{u}{\sqrt{u^2+1}}\:\Theta(u). 
\label{eq:A2}
\end{eqnarray}
Here $F(x)$ and $G(x)$ are arbitrary functions of $x$, $Z$ $(=N-1)$ 
denotes the coordination number, $I_i(t)$ an input signal 
from external sources, $w$ the coupling strength, and $\Theta(u)$
the Heaviside function: $\Theta(u)=1$ for $u > 0$ and
$\Theta(u)=0$ otherwise.
We have included additive and multiplicative noise by $\xi_i(t)$ 
and $\eta_i(t)$, respectively, expressing zero-mean Gaussian white
noise with correlations given by
\begin{eqnarray}
\left< \eta_{i}(t)\:\eta_{j}(t') \right> 
&=& \alpha^2 [\delta_{ij}+c_M(1-\delta_{ij})]\delta(t-t'), 
\label{eq:A3} \\
\left< \xi_{i}(t)\:\xi_{j}(t') \right> 
&=& \beta^2 [\delta_{ij}+c_A(1-\delta_{ij})]]\delta(t-t'), 
\label{eq:A4} \\
\left< \eta_{i}(t)\:\xi_{j}(t') \right> &=& 0,
\label{eq:A5}
\end{eqnarray}
where the bracket $\left< \cdot \right>$ denotes the average,
$\alpha$ ($\beta$) expresses the magnitude of multiplicative 
(additive) noise, and $c_M$ ($c_A$) stands for the degree of 
the spatial correlation in multiplicative (additive) noise. 
Although our results to be present in the following are valid 
for any choice of $H(x)$, we have adopted a simple analytic 
expression given by Eq. (\ref{eq:A2}) in this study.

We assume that external inputs have a variability defined by
\begin{eqnarray}
I_i(t) &=& I(t) + \delta I_i(t), 
\label{eq:A6}
\end{eqnarray}
with
\begin{eqnarray}
\langle \delta I_i(t) \rangle &=& 0, 
\label{eq:A7} \\
\langle \delta I_i(t) \delta I_j(t') \rangle 
&=& \gamma_I [\delta_{ij}  + S_I (1-\delta_{ij}) ]
\delta(t-t'),
\label{eq:A8}
\end{eqnarray}
where $\gamma_I$ and $S_I$ denote the variance and degree of 
the spatial correlation, respectively, in external signals. 
We will investigate the response of the coupled Langevin model
to correlated external inputs given by Eqs. (\ref{eq:A6})-(\ref{eq:A8}).

\subsection{Augmented moment method}

In the AMM \cite{Hasegawa03a,Hasegawa06}, we define the three quantities of 
$\mu(t)$, $\gamma(t)$, and $\rho(t)$ expressed by
\begin{eqnarray}
\mu(t) &=& \langle X(t) \rangle 
= \frac{1}{N} \sum_i \langle x_i(t) \rangle, 
\label{eq:B1}\\
\gamma(t) &=& \frac{1}{N} \sum_i \langle [x_i(t)-\mu(t)]^2 \rangle,
\label{eq:B2}\\
\rho(t) &=& \langle [X(t)-\mu(t)]^2 \rangle,
\label{eq:B3} 
\end{eqnarray}
where 
$X(t)=(1/N) \sum_i x_i(t)$,
$\mu(t)$ expresses the mean, and $\gamma(t)$ and $\rho(t)$ denote 
fluctuations in local ($x_i$) and global variables ($X$),
respectively.
By using the Fokker-Planck equation (FPE), we obtain equations of motion for
$\mu(t)$, $\gamma(t)$ and $\rho(t)$ which are given by
(argument $t$ is suppressed, details being given in the Appendix A)
\begin{eqnarray}
\frac{d \mu}{dt}&=& f_{0}  + h_{0} + f_2 \gamma 
+ \left( \frac{\alpha^2}{2}\right)
[g_{0}g_{1}+3(g_{1}g_{2}+g_{0}g_{3})\gamma], 
\label{eq:C1}\\
\frac{d \gamma}{dt} &=& 2f_{1} \gamma
+  \frac{2 h_{1} w}{Z}  \left(N \rho-\gamma  \right)
+ 2(g_{1}^2+2 g_{0}g_{2})\alpha^2\gamma
+ \gamma_I+\beta^2 + \alpha^2 g_{0}^2, 
\label{eq:C2}\\
\frac{d \rho}{dt} &=& 2 f_{1} \rho 
+ 2 h_1 w \rho
+ 2 (g_{1}^2+2 g_{0}g_{2}) \alpha^2 \rho \nonumber \\
&+& \frac{1}{N}(\gamma_I+\beta^2+\alpha^2 g_0^2)
+\frac{Z}{N}(S_I \gamma_I+ c_A \beta^2 +c_M \alpha^2 g_0^2), 
\label{eq:C3}
\end{eqnarray}
where $f_{\ell}=(1/\ell !)
(\partial^{\ell} F(\mu)/\partial x^{\ell})$,
$g_{\ell}=(1/\ell !)
(\partial^{\ell} G(\mu)/\partial x^{\ell})$, 
$h_{\ell}=(1/\ell !) 
(\partial^{\ell} H(u)/\partial u^{\ell})$ and
$u=w \mu + I$.
%
Original $N$-dimensional stochastic differential equations (DEs) 
given by Eqs. (\ref{eq:A1})-(\ref{eq:A2}) are transformed to 
the three-dimensional deterministic DEs given 
by Eqs. (\ref{eq:C1})-(\ref{eq:C3}). 
For $\gamma_I=S_I=c_A=c_M=0$, equations of motion given by 
Eqs. (\ref{eq:C1})-(\ref{eq:C3}) reduce to those obtained 
in our previous study \cite{Hasegawa07a}.

When we adopt $F(x)$ and $G(x)$ given by
\begin{eqnarray}
F(x)&=&-\lambda x, 
\label{eq:D1} \\
G(x)&=& x,
\label{eq:D2}
\end{eqnarray}
Eqs. (\ref{eq:C1})-(\ref{eq:C3}) become
\begin{eqnarray}
\frac{d \mu}{dt}&=&-\lambda \mu + h_0
+ \frac{\alpha^2 \mu}{2}, 
\label{eq:D3}
\\
\frac{d \gamma}{dt} &=& -2 \lambda \gamma 
+ \frac{2 h_1 w N}{Z} \left( \rho-\frac{\gamma}{N} \right)
+ 2 \alpha^2 \gamma+ P, 
\label{eq:D4}
\\
\frac{d \rho}{dt} &=& - 2 \lambda \rho 
+ 2 h_{1} w \rho 
+ 2 \alpha^2 \:\rho +\frac{(P+ZR)}{N}, 
\label{eq:D5}
\end{eqnarray}
with
\begin{eqnarray}
P &=& \gamma_I+\beta^2+\alpha^2\mu^2, 
\label{eq:F4}\\
R &=& S_I \gamma_I+c_A \beta^2+ c_M \alpha^2 \mu^2,
\label{eq:F5}
\end{eqnarray}
where $h_0=H(w \mu+I)$, and
$P$ and $R$ express uncorrelated and correlated contributions, respectively.
We employ Eqs. (\ref{eq:D3})-(\ref{eq:F5}) in the remainder of this paper.

\subsection{Synchrony and variability}

\subsubsection{Synchrony}

In order to quantitatively discuss the synchronization, we first 
consider the quantity $S'(t)$ given by
\begin{equation}
S'(t)=\frac{1}{N^2} \sum_{i j}<[x_{i}(t)-x_{j}(t)]^2>
=2 [\gamma(t)-\rho(t)].
\label{eq:E1}
\end{equation}
When all neurons are in the same state: $x_{i}(t)=X(t)$ for all $i$
(the completely synchronous state), we obtain $S'(t)=0$ in Eq. (\ref{eq:E1}).
On the contrary, in the asynchronous state where $\rho=\gamma/N$,
it is given by $S'(t)=2(1-1/N)\gamma(t) \equiv S'_{0}(t)$
\cite{Hasegawa07a,Hasegawa03a}.
We may define the normalized ratio for the synchrony given by 
\cite{Hasegawa03a,Hasegawa06}
\begin{equation}
S(t) \equiv 1-\frac{S'(t)}{S'_{0}(t)}
= \left( \frac{N}{Z}\right) 
\left(\frac{\rho(t)}{\gamma(t)}-\frac{1}{N} \right),
\label{eq:E2}
\end{equation}
which is 0 and 1 for completely asynchronous ($S'=S'_0$) and synchronous 
states ($S'=0$), respectively.

\subsubsection{Variability}

The local variability is conventionally given by
\begin{eqnarray}
C_{V}(t) &=& \frac{\sqrt{\langle [\delta x_i(t)]^2 \rangle}}{\mu(t)}
= \frac{\sqrt{\gamma(t)}}{\mu(t)},
\label{eq:E3}
\end{eqnarray}
where $\delta x_i(t)=x_i(t)-\mu(t)$.
Similarly, the global variability is defined by
\begin{eqnarray}
D_V(t) &=&  \frac{\sqrt{\langle [\delta X(t)]^2 \rangle }}{\mu(t)}
= \frac{\sqrt{\rho(t)}}{\mu(t)}
=C_V(t)\;\sqrt{\frac{\rho(t)}{\gamma(t)}},
\label{eq:E4} 
\end{eqnarray}
where $\delta X(t)=X(t)-\mu(t)$.

\subsection{Stationary properties}

The stationary solution of Eqs. (\ref{eq:D3})-(\ref{eq:D5}) is given by
\begin{eqnarray}
\mu &=& \frac{h_0}{(\lambda-\alpha^2/2)}, 
\label{eq:F1}\\
\gamma &=& \frac{P}{2(\lambda-\alpha^2+h_1 w/Z)}
+ \frac{(h_1 w/Z)(P + Z R)}
{2(\lambda-\alpha^2+h_1 w/Z)(\lambda-\alpha^2-h_1 w)}, 
\label{eq:F2}\\
\rho &=& \frac{P + Z R}
{2N(\lambda-\alpha^2-h_1 w)}, 
\label{eq:F3}
\end{eqnarray}
where $\mu$ in $P$ and $R$ of Eqs. (\ref{eq:F4}) and (\ref{eq:F5}) 
is given by Eq. (\ref{eq:F1}). 
We note in Eq. (\ref{eq:F1}) that $\mu$ is increased as $I$ is increased
with an enhancement factor of $1/(\lambda-\alpha^2/2)$.
A local fluctuation $\gamma$ is increased with increasing 
input fluctuations 
($\gamma_I$) and/or noise ($\alpha$, $\beta$) as Eq. (\ref{eq:F2}) shows.
In the limit of $S_I=c_A=c_M=R=w=0.0$, Eqs. (\ref{eq:F2}) and (\ref{eq:F3}) 
lead to $\rho/\gamma=1/N$ which expresses the central-limit theorem.
From Eqs. (\ref{eq:E2}), (\ref{eq:F2}) and (\ref{eq:F3}), we obtain 
\begin{eqnarray}
S &=& \frac{h_1 w P + Z(\lambda -\alpha^2)R}
{P[Z(\lambda-\alpha^2)-h_1 w(Z-1)]+h_1 w Z R},
\label{eq:F6} \\
&=& \frac{h_1 w}{Z(\lambda-\alpha^2)-h_1 w (Z-1)}
\hspace{0.5cm}\mbox{for $S_I=c_A=c_M=0$},
\label{eq:F7}\\
&=& \frac{S_I \gamma_I+c_A \beta^2+ c_M \alpha^2 \mu^2}
{\gamma_I+\beta^2+\alpha^2 \mu^2}
\hspace{1cm}\mbox{for $w=0$},
\label{eq:F8}
\end{eqnarray}
where $P$ and $R$ in Eq. (\ref{eq:F6}) are given by Eqs. (\ref{eq:F4})
and (\ref{eq:F5}), respectively. Equation (\ref{eq:F6}) shows that
the synchrony $S$ is increased with increasing spatial correlations
and/or the coupling. This is more clearly seen in the limit of 
no spatial correlations [Eq. (\ref{eq:F7})] or no couplings [Eq. (\ref{eq:F8})].
The local and global variabilities, $C_V$ and $D_V$, defined by
Eqs. (\ref{eq:E3}) and (\ref{eq:E4}), respectively,
are generally expressed in terms of $P$ and $R$,
and they are given for $w=0.0$ by
\begin{eqnarray}
C_V &=& 
\frac{1}{\mu} \left( \frac{\gamma_I+\beta^2+\alpha^2\mu^2}
{2(\lambda -\alpha^2)} \right)^{1/2}
\hspace{1cm}\mbox{for $w=0$}, \label{eq:F17} \\
D_V &=& C_V 
%
\left( \frac{(1+ZS_I)\gamma_I+(1+Z c_A)\beta^2+(1+Z c_M)\alpha^2\mu^2}
{N(\gamma_I+\beta^2+\alpha^2\mu^2)} \right)^{1/2}
\label{eq:F14} \nonumber \\
&&\hspace{6cm}\mbox{for $w=0$}.
\label{eq:F18}
\end{eqnarray}
The local variability $C_V$ 
only weakly depends on the spatial correlation through the coupling, and 
it is independent of the correlation for $w=0$. In contrast, the global 
variability $D_V$ is increased (decreased) for positive (negative) correlations.
In the limit of $S_I=c_A=c_M=w=0.0$, Eq. (\ref{eq:F18}) yields 
$D_V = C_V/\sqrt{N}$ expressing a smaller global variability
in a larger-$N$ ensemble (the population or pooling effect)
\cite{Averbeck06,Liu01,Wang04,Kwon05}.

The stability condition around the stationary state given by 
Eqs. (\ref{eq:F1})-(\ref{eq:F3}) may be examined from eigenvalues of 
the Jacobian matrix of Eqs. (\ref{eq:D3})-(\ref{eq:D5}), which are 
given by 
\begin{eqnarray}
\lambda_1 &=& -\lambda+\frac{\alpha^2}{2}+ h_1 w, \\
\lambda_2 &=& -2 \lambda+2 \alpha^2-\frac{2 h_1 w}{Z}, \\
\lambda_3 &=& -2 \lambda+2 \alpha^2+2 h_1 w.
\label{eq:F9} 
\end{eqnarray}
The first eigenvalue of $\lambda_1$ arises from an equation of 
motion for $\mu$, which is decoupled from the rest of variables. 
The stability condition for $\mu$ is given by
\begin{equation}
h_1 w < (\lambda-\alpha^2/2).
\label{eq:F10}
\end{equation}
The stability condition for $\gamma$ and $\rho$ is given by
\begin{equation}
-Z (\lambda-\alpha^2) < h_1 w < (\lambda-\alpha^2).
\label{eq:F11}
\end{equation}
Then for $\lambda-\alpha^2 < h_1 w < \lambda -\alpha^2/2$,
$\gamma$ and $\rho$ are unstable but $\mu$ remains stable. 

It is note that there is a limitation in a parameter value of $c$, 
as given by
\begin{eqnarray}
-\frac{1}{Z} & \leq &
\left( \frac{S_I \gamma_I+c_A \beta^2+ c_M \alpha^2 \mu^2}
{\gamma_I+\beta^2+\alpha^2\mu^2}\right) \leq 1, 
\label{eq:F13}
\end{eqnarray}
which arises from the condition given by $0 \leq \rho \leq \gamma$
[see Eqs. (\ref{eq:B3}) and (\ref{eq:E1})]. When $\beta=\gamma_I=0$, 
for example, a physically conceivable value of $c_M$ is given 
by $-1/Z \leq c_M \leq 1$.

$c_M$ dependence of the synchrony $S$ is shown in Fig. \ref{figA} 
where $\alpha=0.1$, $\beta=0.1$, $c_A=0.1$, $\gamma_I=S_I=0$ and $N=100$.
The condition given by Eq. (\ref{eq:F13}) yields that $-0.44 < c_M < 1.0 $ 
with $\mu=0.5$ for a given set of parameters.
We note that the synchrony is increased with increasing $s$, and that
the effect of the correlated variability is more considerable 
for larger $\mu$ and $w$, as Eq. (\ref{eq:F6}) shows.

\subsection{Dynamical properties}

In order to study the dynamical properties of our model 
given by Eqs. (\ref{eq:A1})-(\ref{eq:A2}), we have performed direct simulations
(DSs) by using the Heun method \cite{Heun,Heun2} with a time step of 0.0001:
DS results are averages of 100 trials. AMM calculations have been performed 
for Eqs. (\ref{eq:D3})-(\ref{eq:D5}) by using the fourth-order Runge-Kutta method 
with a time step of 0.01. We consider a set of typical parameters of
$\lambda=1.0$, $\alpha=0.1$, $\beta=0.1$, $w=0.5$, $\gamma_I=S_I=0$ and $N=100$.
We apply a pulse input given by
\begin{eqnarray}
I(t) &=& A \:\Theta(t-40) \Theta(60-t)+ A_b,
\label{eq:H1}
\end{eqnarray}
with $A=0.4$, $A_{b}=0.1$, where $\Theta(x)$ denotes the Heaviside function:
$\Theta(x)=1$ for $x \geq 0$ and 0 otherwise.

Figures \ref{figB}(a)-\ref{figB}(d) show time courses of $\mu(t)$, $\gamma(t)$, 
$S(t)$ and $C_{V}(t)$ for the correlated multiplicative noise 
($c_A=0.0$ and $c_M=0.5$). $\mu(t)$ and $S(t)$ are increased by
an applied input at $40 \leq t < 60$ shown by the chain curve
in Fig. \ref{figB}(a), by which $\gamma(t)$ is slightly increased.
The variability $C_V(t)$ is decreased because of an increased $\mu(t)$.
The results of the AMM shown by the solid curves are in fairly good 
agreement with those of DS shown by the dashed curves.

In contrast, Figs. \ref{figC}(a)-\ref{figC}(d) show time courses 
of $\mu(t)$, $\gamma(t)$, 
$S(t)$ and $C_{V}(t)$ for the correlated additive noise ($c_A=0.1$ and $c_M=0.0$).
With an applied pulse input, $\mu(t)$ is increased
and $\gamma(t)$ is a little increased, as in the case of 
Figs. \ref{figB}(a) and \ref{figB}(b).
However, the synchrony $S(t)$ is decreased in Fig. \ref{figC}(c)
while it is increased in Fig. \ref{figB}(c). 
This difference arises from the fact that a decrease in $S(t)$ 
in the former case is mainly due to an increase in $P$ of the denominator 
of Eq. (\ref{eq:F6}), while in the latter case, its increase arises
from an increase in $R$ of the numerator of Eq. (\ref{eq:F6}).
This point is more easily realized for $w=0$, for which Eq. (\ref{eq:F8}) 
yields 
\begin{eqnarray}
S &=& \frac{c_M \alpha^2 \mu^2}{\beta^2+\alpha^2 \mu^2}
\hspace{1cm} \mbox{for $c_A=0$}, \\
&=& \frac{c_A \beta^2}{\beta^2+\alpha^2 \mu^2}
\hspace{1cm} \mbox{for $c_M=0$}.
\end{eqnarray}
The situation is almost the same even for finite $w$, as Figs. \ref{figB}(c) 
and \ref{figC}(c) show. In both Figs. \ref{figB}(d) and \ref{figC}(d), 
$C_{V}(t)$ is decreased by an applied input because of an increased $\mu(t)$. 

\section{Discussion}

\subsection{A comparison with related studies}

We have investigated the stationary and dynamical properties of
the spatially correlated Langevin model given by 
Eqs. (\ref{eq:A1})-(\ref{eq:A2}). In Ref. \cite{Hasegawa08a}, 
we discussed the Fisher information in the Langevin model subjected to
uncorrelated additive and multiplicative noise, which is a typical 
microscopic model showing the nonextensive behavior \cite{Tsallis88}. 
It is interesting to calculate the Fisher information
of the Langevin model with correlated multiplicative noise.
Such a calculation needs to solve the FPE of 
the Langevin model given by Eq. (\ref{eq:A9})
because the Fisher information is expressed in terms of derivatives of 
the probability distribution.
For additive noise only ($\alpha=c_M=0$), the stationary probability 
distribution $p(\{ x_k \})$ 
is expressed by the multivariate Gaussian distribution given by
\begin{eqnarray}
p(\{ x_k \}) &\propto&
\exp\left[-\frac{1}{2} \sum_{ij} (x_i-\mu)({\sf Q}^{-1})_{ij}(x_j-\mu)\right],
\label{eq:L1}
\end{eqnarray}
with the covariance matrix, ${\sf Q}$, expressed by
\begin{eqnarray}
Q_{ij} &=& \sigma^2 [\delta_{ij}+ c_A (1-\delta_{ij})],
\label{eq:L2}
\end{eqnarray}
where $\mu=H/\lambda$ and $\sigma^2=\beta^2/2 \lambda$.
From Eqs. (\ref{eq:L1}) and (\ref{eq:L2}), 
we obtain the Fisher information given by \cite{Abbott99}
\begin{eqnarray}
g &=& \frac{N}{\sigma^2 [1+(N-1) c_A]}.
\label{eq:L3}
\end{eqnarray}

When multiplicative noise exists, a calculation 
of even stationary distribution becomes difficult, and it is generally 
not given by the Gaussian. The stationary distribution 
for {\it uncorrelated} additive and multiplicative noise ($G(x)=x$, 
$c_A=c_M=\gamma_I=0.0$) is given by 
\cite{Hasegawa07a,Hasegawa08a,Sakaguchi01,Anten02}
\begin{eqnarray}
p(\{ x_k \}) &\propto& \prod_i
\left(\beta^2+ \alpha^2 x_i^2 \right)^{-(\alpha^2/\lambda+1/2)} 
e^{(2 H/\alpha \beta)
\tan^{-1} ( \alpha x_i/\beta )}.
\label{eq:L4}
\end{eqnarray}
In the limit of $\alpha=0.0$ and $\beta \neq 0.0$ ({\it i.e.} uncorrelated additive
noise only), Eq. (\ref{eq:L4}) becomes the
Gaussian distribution given by
\begin{eqnarray}
p(\{x_k \}) & \propto & \prod_i e^{-(\lambda/\beta^2)(x_i-\mu)^2}.
\end{eqnarray}
In the opposite limit of $\alpha \neq 0.0$, $\beta=0.0$ and $H > 0$
({\it i.e.} uncorrelated multiplicative noise only), 
Eq. (\ref{eq:L4}) reduces to
\begin{eqnarray}
p(\{x_k \}) & \propto & \prod_i \;x_i^{-(2\alpha^2/\lambda+1)}
e^{-2 H/\alpha^2 x_i} \; \Theta(x_i), 
\label{eq:L8}  
\end{eqnarray}
yielding the Fisher information given by 
\begin{eqnarray}
g &=& \frac{N q^4}{\sigma_q^2}
= \frac{2 N \lambda q^4}{\alpha^2 \mu^2},
\label{eq:L9} 
\end{eqnarray}
where $q=(2\lambda+3\alpha^2)/(2\lambda+\alpha^2)$
and $\sigma_q^2=\alpha^2 \mu^2/2 \lambda $ \cite{Hasegawa08a}. 

Unfortunately, we have not succeeded in obtaining the analytic expression 
for the stationary distribution 
of the Langevin model including {\it correlated} multiplicative noise.
In some previous studies \cite{Abbott99}-\cite{Wu04}, the stationary 
distribution for correlated multiplicative
noise only ($G(x)=x$, $c_A=\beta=0.0$ and $H=\mu$) is assumed to be 
expressed by the Gaussian distribution given by Eq. (\ref{eq:L1}) 
with the covariance matrix given by
\begin{eqnarray}
Q_{ij} &=& \sigma_M^2 \:\mu_i \mu_j [\delta_{ij}+ c_M (1-\delta_{ij})],
\label{eq:L6}
\end{eqnarray}
where $\mu_i$ ($=\langle x_i \rangle$) denotes the average of $x_i$ 
and $\sigma_M^2$ a variance due to multiplicative noise.
This is equivalent to assume that the multiplicative-noise
term in the FPE given by Eq. (\ref{eq:A9}) is approximated as
\begin{eqnarray}
&& \frac{\alpha^2}{2} \sum_{i} \sum_j
[\delta_{ij}+c_M(1-\delta_{ij})]
\frac{\partial}{\partial x_{i}} 
x_i \frac{\partial}{\partial x_j}
x_j \:p(\{ x_{k} \}),  \nonumber \\ 
&\simeq& \frac{\alpha^2}{2} 
\sum_{i} \sum_j 
[\delta_{ij}+c_M(1-\delta_{ij})]
\frac{\partial}{\partial x_{i}} 
\langle x_i \rangle
\frac{\partial}{\partial x_j}
\langle x_j \rangle \:p( \{ x_k \}),\nonumber \\
&=& \frac{\alpha^2}{2}
\sum_j \sum_{i}  \mu_i \mu_j 
[\delta_{ij}+c_M(1-\delta_{ij})]
\frac{\partial}{\partial x_{i}}  
\frac{\partial}{\partial x_j} p(\{ x_{k} \}).
\label{eq:L7} 
\end{eqnarray}
If we adopt the Gaussian approximation given by Eq. (\ref{eq:L7}),
with which multiplicative noise may be treated in the same way as
additive noise, we obtain the AMM equations given by 
Eqs. (\ref{eq:D3})-(\ref{eq:D5}) but without the third term of 
$\alpha^2 \mu/2$ in Eq. (\ref{eq:D3}).

By using the Gaussian approximation given by Eq. (\ref{eq:L6}),
Abbott and Dayan (AD) \cite{Abbott99} obtained the Fisher information 
expressed by [Eq. (4.7) of Ref. \cite{Abbott99}]
\begin{eqnarray}
g_{AD} &=& \frac{N K}{\sigma_M^2 [1+(N-1)c_M]}
+ 2 N K, \label{eq:H5} \\
&=& \frac{1}{ \sigma_M^2 \mu^2 c_M}+
\frac{2N}{\mu^2}
\hspace{3.6cm}\mbox{for $N \rightarrow \infty $}, 
\label{eq:H6} \\
&=& \frac{N}{\sigma_M^2 \mu^2} +\frac{2N}{\mu^2}
\hspace{2cm}\mbox{for $c_M=0$},
\label{eq:H4}
\end{eqnarray}
with $K=N^{-1}\sum_i[d \ln H(\mu_i)/d \mu_i]^2=1/\mu^2$. 
Equation (\ref{eq:H4}) is not in agreement with the exact expression 
given by Eq. (\ref{eq:L9}) for uncorrelated multiplicative noise only.

In stead of using the Langevin model, we may alternatively
calculate the Fisher information of 
a spatially correlated nonextensive system by using the maximum-entropy method.
In our recent paper \cite{Hasegawa08b}, we have obtained the analytic, 
stationary probability distribution which maximizes the Tsallis entropy 
\cite{Tsallis88}
under the constraints for a given set of the variance ($\sigma^2$)
and covariance ($c\:\sigma^2$). The Fisher information is expressed by 
\cite{Hasegawa08b}
\begin{eqnarray}
g &=& \frac{N}{\sigma^2[1+(N-1) c]},
\label{eq:F12} \\
&=& \frac{1}{c \:\sigma^2}
\hspace{2cm}\mbox{for $N \rightarrow \infty $}, 
\\
&=& \frac{N}{\sigma^2}
\hspace{2cm}\mbox{for $c = 0.0$}.
\label{eq:F14}
\end{eqnarray}
The Fisher information given by Eq. (\ref{eq:F12}) is
increased (decreased) by a negative (positive) correlation. 
This implies from the Cram\'{e}r-Rao theorem that
an unbiased estimate of fluctuations is improved by a negative spatial 
correlation, by which the synchrony is decreased as shown by Eqs. (\ref{eq:F6}) 
and (\ref{eq:F8}). 
$N$ and $c$ dependences of the Fisher information 
given by Eq. (\ref{eq:F12}) are different from those of $g_{AD}$ given 
by Eq. (\ref{eq:H5}), although they are the same as those 
for additive noise only [Eq. (\ref{eq:L3})]. 
It is noted that the Gaussian approximation given 
by Eq. (\ref{eq:L6}) or (\ref{eq:L7}) assumes the Gaussian distribution, 
although multiplicative noise generally yields the non-Gaussian distribution 
as shown by Eqs. (\ref{eq:L4}) and (\ref{eq:L8}).
The spurious second term ($2NK$) in Eq. (\ref{eq:H5}) 
which is independent of $c_M$ and $\sigma_M^2$, arises from an inappropriate 
Gaussian approximation. In discussing the Fisher information of 
spatially correlated nonextensive systems, we must take into account 
the detailed structure of the non-Gaussian distribution.

\subsection{Application to neuronal ensembles}

When $\gamma_I$ and $S_I$ in Eq. (\ref{eq:A8}) are allowed to be
time dependent, they may carry input information.
This is easily realized if the AMM equations given by 
Eqs. (\ref{eq:D3})-(\ref{eq:D5}) are explicitly expressed 
in terms of $\mu$, $\gamma$, and $S$ as
\begin{eqnarray}
\frac{d \mu}{dt}&=&-\lambda \mu + h_0
+ \frac{\alpha^2 \mu}{2}, 
\label{eq:G3}
\\
\frac{d \gamma}{dt} &=& -2 \lambda \gamma 
+ 2 h_1 w \gamma S
+ 2 \alpha^2 \gamma  + \gamma_I(t) + \alpha^2 \mu^2 + \beta^2, 
\label{eq:G4} \\
\frac{d S}{dt} &=& - \frac{S}{\gamma} 
[\gamma_I(t)+\alpha^2 \mu^2+\beta^2]
+\frac{1}{\gamma}
[\gamma_I(t) S_I(t)+ c_M \alpha^2 \mu^2+c_A \beta^2] \nonumber \\
&+& \left( \frac{2 h_1 w}{Z} \right) (1+ZS)(1-S), 
\label{eq:G5}
\end{eqnarray}
which are derived with the use of Eq. (\ref{eq:E2}).

In order to numerically examine the possibility that input information is 
conveyed by $\gamma_I(t)$ and $S_I(t)$, we first apply a fluctuation-driven 
input given by
\begin{equation}
\gamma_I(t)=  B \:\Theta(t-40) \Theta(60-t)+B_b,
\label{eq:H2}
\end{equation}
with $B=0.4$, $B_b=0.1$, $I(t)=0.1$ and $S_I(t)=0.1$ for $\lambda=1.0$, 
$c_A=0.0$, $c_M=0.5$, $\alpha=0.1$, $\beta=0.1$ and $N=100$.
Time courses of $\mu(t)$, $\gamma(t)$, $S(t)$ and $C_V(t)$
are shown in Figs. \ref{figG}(a)-\ref{figG}(d): 
the chain curves in Figs. \ref{figG}(a) and \ref{figG}(b)
express $I(t)$ and $\gamma_I(t)$, respectively. When the magnitude of 
$\gamma_I(t)$ is increased at $40 \leq t < 60$, $\gamma(t)$ and $C_{V}(t)$ 
are much increased, while there is no changes in $\mu(t)$ because
it is decoupled from the rest of variables in Eq. (\ref{eq:D3}).
$S(t)$ is slightly modified only at $t \sim 40$ and $t \sim 60$
where the $\gamma_I(t)$ is on and off.

Next we apply a synchrony-driven input $S_I(t)$ given by
\begin{equation}
S_I(t)= C \:\Theta(t-40) \Theta(60-t)+C_b,
\label{eq:H3}
\end{equation}
with $C=0.4$, $C_b=0.1$, $I(I)=0.1$ and $\gamma_I(t)=0.1$.
Figures \ref{figH}(a)-\ref{figH}(d) show time courses of $\mu(t)$, 
$\gamma(t)$, $S(t)$ and $C_V(t)$: the chain curves 
in Figs. \ref{figH}(a) and \ref{figH}(c) express $I(t)$ and $S_I(t)$, 
respectively.
An increase in synchrony-driven input at $40 \leq t < 60$
induces a significant increase in $S(t)$ and slight increases in 
$\gamma(t)$ and $C_{V}(t)$, but no changes in $\mu(t)$. 

When we regard a variable $x_i$ in the Langevin model given by
Eqs. (\ref{eq:A1})-(\ref{eq:A2}) as the firing rate $r_i$ ($> 0$) 
of a neuron $i$ in a neuron ensemble, our model expresses the
neuronal model proposed in 
Refs. \cite{Hasegawa07b,Hasegawa07c}. 
It belongs to the firing-rate (rate-code) models such as the 
Wilson-Cowan \cite{Wilson72} and Hopfield models \cite{Hopfield84}, 
in which a neuron is regarded as a transducer from input rate signals 
to output rate ones. Alternative neuronal models are spiking-neuron 
(temporal-code) models such as the Hodgkin-Huxley \cite{Hodgkin52}, 
FitzHugh-Nagumo \cite{FitzHugh61,Nagumo62}
and integrate-and-fire (IF) models \cite{ReviewIF}.
Various attempts have been proposed to obtain the firing-rate model, 
starting from spiking-neuron models \cite{Amit91}-\cite{Oizumi07}.
It is difficult to analytically calculate the firing rate
based on spiking-neuron models, except for the IF-type model
\cite{ReviewIF}. 
It has been shown with the use of the IF model that information transmission 
is possible by noise-coded signals \cite{Lindner01,Renart07}, and
that the modulation of the synchrony is possible without a change 
in firing rate \cite{Heinzle07}.
Model calculations shown in Figs. \ref{figG} and \ref{figH}
have demonstrated the possibility that information may be conveyed 
by $\gamma_I(t)$ and $S_I(t)$, which is partly supported by
results of the IF model \cite{Lindner01}-\cite{Heinzle07}.
Some relevant results have been reported in neuronal experiments
\cite{Riehle97}-\cite{Tiesinga04}.
In motor tasks of monkey, firing rate and synchrony 
are considered to encode
behavioral events and cognitive events, respectively \cite{Riehle97}. 
During visual tasks, rate and synchrony are suggested to
encode task-related signals and expectation, respectively
\cite{Oliveira97}. A change in synchrony may amplify behaviorally 
relevant signals in V4 of monkey \cite{Fries01}. The synchrony is modified 
without a change in firing rate in some experiments
\cite{Riehle97,Fries01,Grammont03}.
Synchrony-dependent firing-rate signal is shown to propagate
in iteratively constructed networks {\it in vitro}
\cite{Reyes03}.

\section{Conclusion}

With the use of DSs and the AMM \cite{Hasegawa03a,Hasegawa06}, the effects of 
spatially correlated additive and multiplicative noise have been discussed
on the synchrony and variability in the coupled Langevin model.
Our calculations have shown the following:
(i) the synchrony is increased (decreased) by the positive (negative) 
correlation in additive and multiplicative noise [Eq. (\ref{eq:F6})],
(ii) although an applied pulse input works to increase the synchrony
for correlated multiplicative noise, it is possible to decrease 
the synchrony when correlated additive noise coexists with
uncorrelated multiplicative one,
(iii) the local variability $C_V$ is almost independent of 
spatial correlations, while global variability $D_V$ is increased
(decreased) with increasing the positive (negative) correlation, and 
(iv) information may be carried by variance and synchrony in input signals.
The item (iv) is consistent with the results of Refs. 
\cite{Lindner01}-\cite{Heinzle07} and elucidates some phenomena observed 
in neuronal experiments \cite{Riehle97}-\cite{Tiesinga04}.

Although we have applied the AMM to the Langevin model in this paper, 
it is possible to apply it to other types of stochastic neuronal models such 
as the FitzHugh and Hodgkin-Huxley models subjected to correlated additive 
and multiplicative noise, which is left as our future study.

\section*{Acknowledgments}
This work is partly supported by
a Grant-in-Aid for Scientific Research from the Japanese 
Ministry of Education, Culture, Sports, Science and Technology.  

\vspace{0.5cm}
\appendix

\noindent
{\large\bf Appendix A: Derivation of the AMM equations}
\renewcommand{\theequation}{A\arabic{equation}}
\setcounter{equation}{0}

The Fokker-Planck equation (FPE) for the Langevin equation 
given by Eqs. (\ref{eq:A1})-(\ref{eq:A2}) in the Stratonovich representation 
is expressed by \cite{Ibanes99,Risken92}
\begin{eqnarray}
&&\frac{\partial}{\partial t}\: p(\{x_{k} \},t)=
-\sum_{i} \frac{\partial}{\partial x_{i}}\{ [F(x_{i}) + H(u_{i})]
\:p(\{ x_{k} \},t)\}  
\nonumber \\
&&+\frac{\gamma_I}{2}\sum_{i}\sum_{j}
\frac{\partial^2}{\partial x_{i} \partial x_{j}} 
\{ [\delta_{ij}+S_I (1-\delta_{ij}) ]
\:p(\{ x_{k} \},t) \}
\nonumber \\
&&+\frac{\beta^2}{2}\sum_{i}\sum_{j}
\frac{\partial^2}{\partial x_{i} \partial x_{j}} 
\{ [\delta_{ij}+c_A (1-\delta_{ij}) ]
\:p(\{ x_{k} \},t) \}
\nonumber \\
&&+\frac{\alpha^2}{2}\sum_{i} \sum_j
[\delta_{ij}+c_M(1-\delta_{ij})]
\frac{\partial}{\partial x_{i}} 
\{ G(x_i) 
\frac{\partial}{\partial x_j}
[ G(x_{j}) \:p(\{ x_{k} \},t) ] \}, 
\label{eq:A9}
\end{eqnarray}
where $u_i=(w/Z) \sum_{j(\neq i)} x_j+I$ and $H'(u_i)$ is absorbed 
in a new definition of $\gamma_I$ in its second term.

Equations of motion for moments, $\langle x_{i}  \rangle$
and $\langle x_{i} \:x_{j} \rangle$, are derived with the use of FPE
\cite{Hasegawa07a}:
\begin{eqnarray}
\frac{d \langle x_i \rangle}{dt}
&=& \langle F(x_i) + H(u_i) \rangle
+\frac{\:\alpha^2}{2} \langle G'(x_i)G(x_i) \rangle,
\label{eq:A10}
\\
\frac{d \langle x_i \:x_j \rangle}{dt}
&=& \langle x_i\:[F(x_j)+H(u_j)] \rangle 
+ \langle x_j\: [F(x_i)+ H(u_i)] \rangle \nonumber \\
&+& \frac{\alpha^2}{2}
[\langle x_i G'(x_j) G(x_j) \rangle
+ \langle x_j G'(x_i) G(x_i)\rangle] \nonumber \\
&+& \delta_{ij} [\gamma_I+\beta^2 +\alpha^2\:\langle G(x_i)^2 \rangle ]
\nonumber \\
&+& (1-\delta_{ij}) 
[S_I\gamma_I+c_A \beta^2 + c_M \alpha^2 \langle G(x_i)G(x_j) \rangle].
\label{eq:A11}
\end{eqnarray}

In the AMM \cite{Hasegawa03a,Hasegawa06}, the three quantities of 
$\mu$, $\gamma$ and $\rho$ are defined by Eqs. (\ref{eq:B1})-(\ref{eq:B3}).
We use the expansion given by
\begin{eqnarray}
x_i &=& \mu + \delta x_i, 
\label{eq:B4}
\end{eqnarray}
and the relations given by
\begin{eqnarray}
\frac{d \mu}{dt} &=& \frac{1}{N} \sum_i 
\frac{d \langle r_i \rangle}{dt}, 
\label{eq:B5}
\\
\frac{d \gamma}{dt} &=& \frac{1}{N} \sum_i 
\frac{d \langle (\delta r_i)^2 \rangle}{dt},
\label{eq:B6} 
\\
\frac{d \rho}{dt} &=& \frac{1}{N^2} \sum_i \sum_j 
\frac{d \langle \delta r_i \delta r_j \rangle}{dt}.
\label{eq:B7}
\end{eqnarray}
For example, Eq. (\ref{eq:B5}) for $d \mu/dt$ is calculated as follows:
\begin{eqnarray}
\frac{1}{N} \sum_i \langle F(r_i) \rangle
&=& f_0 +f_2 \gamma,
\label{eq:B8}\\
\frac{1}{N} \sum_i \langle H(u_i) \rangle
&=& h_0,
\label{eq:X10}
\\
\frac{1}{N} \sum_i \langle G'(r_i) G(r_i) \rangle
&=& g_0 g_1 + 3(g_0 g_3+g_1 g_2) \gamma.
\label{eq:B9}
\end{eqnarray}
Equations (\ref{eq:B6}) and (\ref{eq:B7}) are calculated in a similar way.
Then, we have obtained equations of motion for $\mu(t)$, 
$\gamma(t)$ and $\rho(t)$ given by Eqs. (\ref{eq:C1})-(\ref{eq:C3}).


\newpage

\begin{figure}
\begin{center}
\end{center}
\caption{
(Color online)
$c_M$ dependence of the stationary synchrony
$S$ for $w=0.5$ (solid curves) and $w=0.0$ (dashed curves)
with $\alpha=0.1$, $\beta=0.1$, $c_A=0.1$, $\gamma_I=S_I=0$ and $N=100$,
$\mu$ being treated as a parameter.
}
\label{figA}
\end{figure}

\begin{figure}
\begin{center}
\end{center}
\caption{
(Color online)
Time courses of (a) $\mu(t)$, (b) $\gamma(t)$, (c) $S(t)$ and (d) $C_{V}(t)$ 
with the correlated multiplicative noise 
($c_A=0.0$, $c_M=0.5$, $\alpha=0.1$, $\beta=0.1$) for a pulse input
given by Eq. (\ref{eq:H1}) with $A=0.4$, $A_{b}=0.1$:
the solid and dotted curves express results of the AMM and DS, respectively: 
the chain curve in (a) expresses an input of $I(t)$
($\lambda=1.0$, $S_I=\gamma_I=0.0$ and $N=100$).
}
\label{figB}
\end{figure}

\begin{figure}
\begin{center}
\end{center}
\caption{
(Color online)
Time courses of (a) $\mu(t)$, (b) $\gamma(t)$, (c) $S(t)$ and (d) $C_{V}(t)$ 
with the correlated additive noise
($c_A=0.1$, $c_M=0.0$, $\alpha=0.1$, $\beta=0.1$) for a pulse input
given by Eq. (\ref{eq:H1}) with $A=0.4$ and $A_{b}=0.1$:
the solid and dashed curves denote results of the AMM and DSs, respectively:
the chain curve in (a) expresses an input of $I(t)$
($\lambda=1.0$, $S_I=\gamma_I=0.0$ and $N=100$).
}
\label{figC}
\end{figure}

\begin{figure}
\begin{center}
\end{center}
\caption{
(Color online)
Time courses of (a) $\mu(t)$, (b) $\gamma(t)$, (c) $S(t)$ and (d) $C_{V}(t)$ 
for a fluctuation-driven input of $\gamma_I(t)$ given by Eq. (\ref{eq:H2}) with 
$S_I=0.1$, $B=0.4$ and $B_{b}=0.1$: the solid and dotted curves express results of 
the AMM and DS, respectively: the chain curves in (a) and (b) 
express inputs of $I(t)$ and $\gamma_I(t)$, respectively
($\lambda=1.0$, $c_A=0.0$, $c_M=0.5$, $\alpha=0.1$, $\beta=0.1$ and $N=100$).
}
\label{figG}
\end{figure}

\begin{figure}
\begin{center}
\end{center}
\caption{
(Color online)
Time courses of (a) $\mu(t)$, (b) $\gamma(t)$, (c) $S(t)$ and (d) $C_{V}(t)$ 
for a synchrony-driven input of $S_I(t)$ given by Eq. (\ref{eq:H3}) with 
$\gamma_I=0.1$, $C=0.4$ and $C_{b}=0.1$: the solid and dotted curves express results 
of the AMM and DS, respectively: the chain curves in (a) and (c) express
inputs of $I(t)$ and $S_I(t)$, respectively
($\lambda=1.0$, $c_A=0.0$, $c_M=0.5$, $\alpha=0.1$, $\beta=0.1$ and $N=100$).
}
\label{figH}
\end{figure}


\begin{thebibliography}{99}
\bibitem{Heun}
In the Heun method for the ordinary differential equation of 
$dx/dt=f(x)$,
a value of $x$ at $t+h$ is evaluated by
$x(t+h)=x(t)+(h/2)[f(x_0)+f(x_1)]$ with $x_0=x(t)$
and $x_1=x(t)+h f(x(t))$, while it is given by
$x(t+h)=x(t)+ h f(x(t))$ in the Euler method,
$h$ being the time step.
The Heum method for the stochastic ordinary differential equation
meets the Stratonovich calculus employed in the FPE 
\cite{Heun2}.
\end{thebibliography}
\end{document}